\documentclass[prl,twocolumn,floats, superscriptaddress,preprintnumbers, nofootinbib]{revtex4}

\usepackage{hyperref, subfigure}

\usepackage{color}
\usepackage{amsmath,amsfonts,amssymb, graphicx}
\newcommand{\be}{\begin{equation}}
\newcommand{\ee}{\end{equation}}
\newcommand{\bea}{\begin{eqnarray}}
\newcommand{\eea}{\end{eqnarray}}

\begin{document}
            
\preprint{CERN-PH-TH/2009-111}            
                                                                               
\title{Nuclear Physics from lattice QCD at strong coupling}

\author{Ph.~de Forcrand}
\affiliation{Institute for Theoretical Physics, ETH Z\"urich, CH-8093 Z\"urich, Switzerland}
\affiliation{CERN, Physics Department, TH Unit, CH-1211 Geneva 23, Switzerland}
\author{M.~Fromm}
\affiliation{Institute for Theoretical Physics, ETH Z\"urich, CH-8093 Z\"urich, Switzerland}

\begin{abstract}
We study numerically the strong coupling limit of lattice QCD with one
flavor of massless staggered quarks. We determine the complete phase diagram
as a function of temperature and chemical potential, including a tricritical point. We clarify the nature of the low temperature dense phase, which is strongly bound ``nuclear'' matter. This strong binding is explained by the nuclear potential, which we measure. Finally, we determine, from this first-principle limiting case of QCD, the masses of ``atomic nuclei'' up to $A=12$ ``carbon''.
\end{abstract}

\maketitle


It has been a long-standing goal to determine the properties of nuclear matter
from first principles, using the theory which describes the 
interactions inside each nucleon, namely quantum chromodynamics (QCD).
Lattice QCD simulations have been extremely successful at determining the
properties of a single proton, neutron or other hadron~\cite{Durr2008}, and
should in principle be appropriate for studying nuclear matter as well. Great 
progress has been made in this area, with the study of 
nuclear scattering lengths and potentials~\cite{Beane_Savage, Ishii:2008, Gardestig:2009}, and of the energies of
two- and three-baryon systems~\cite{Beane:2009}. Nevertheless,
in these pio\-neering studies the quark masses are still far 
from their real-world values and the matter density is still very small. 
The {\it ab initio} study of real-world nuclear
matter remains a distant goal. The technical issue of simulating 
quarks which are as light as in nature is being resolved~\cite{Jansen2008}, but two more
obstacles stand in the way. $(i)$ There is a large scale separation ${\cal O}(50)$ between the nucleons'
masses and their binding energy. Accuracy on the latter requires excellent
control over errors. $(ii)$ There is a severe ``sign problem'': the functional integral
at nonzero nuclear density has an oscillatory integrand,
which causes large numerical cancellations and altogether prevents its 
usual interpretation as a Monte Carlo sampling probability.
For these two reasons, the state-of-the-art approach is that of effective
field theory~\cite{nuclearEFT}, whose couplings still need to be
derived by matching to QCD.\\
\indent Here, we consider a limit of lattice QCD where the above obstacles can be 
overcome: the limit of infinite coupling. Because of asymptotic freedom,
the lattice spacing $a$ goes to zero with the bare coupling $g$ as
$a \propto \exp(-\frac{4\pi^2}{33} \beta)$ with $\beta\!=\! 6/g^2$. In the opposite limit $\beta=0$ we will clearly have large lattice artifacts. Moreover, different discretizations of continuum QCD, all equivalent at weak coupling
$\beta \!\gg\! 1$, may behave differently. We choose the staggered discretization
of the Dirac operator and study the Euclidean partition function
\vspace{-2mm}\be
	Z (m_q, \mu) = \int\mathcal{D}U\mathcal{D}\bar\chi\mathcal{D}\chi ~ \mathrm{e}^{S_\mathrm{F}}\,,
\label{eq:Z}
\ee
\vspace{-1mm}
with gauge links $U$ in $SU(3)$ and action
\be
	S_\mathrm{F} = \!\!\!\! \sum_{x,\nu=1,4} \!\!\! \eta_{x,\hat{\nu}}\bar\chi_x\left[U_{x,\hat\nu}\chi_{x+\hat\nu} - U^{\dagger}_{x-\hat\nu,\hat\nu}\chi_{x-\hat\nu}\right] + 2m_q\sum_x\bar\chi_x\chi_x\label{sc_action}
\ee
on a four-dimensional $N_s^3\times N_\tau$ lattice, with antiperiodic boundary
conditions in Euclidean time for the fermions $\chi$, periodic otherwise.
The $\eta_{x,\hat\nu} \!=\! (-1)^{\sum_{\rho<\nu}x_\rho}$, $\eta_{x,\hat1} \!=\! 1$ are the usual staggered phases.
The chemical potential $\mu$ and an anisotropy $\gamma$ are introduced by multiplying the time-like
gauge links $U_{x,\pm\hat{4}}$ by $\gamma\exp(\pm a\mu/\gamma)$ in the forward and backward directions, respectively.
The usual plaquette term which accounts for the gluonic action is absent here,
since it is multiplied by $\beta$. The anisotropy $\gamma$
allows the temperature $T$ to be varied continuously, via $T \!=\! a^{-1} \gamma^2/N_\tau$ at infinite coupling~\cite{Damgaard}. In this Letter, we \nolinebreak consider \linebreak only one quark species and set its mass $m_q$ to zero.

Our model is clearly very far from continuum QCD which contains two light
quark flavors: our ``quarks'' live on a coarse cubic crystal, come in only 
one flavor and have no spin. 
This causes considerable changes in the symmetries of the theory, its 
bound states and their interactions, and its phase diagram.
Our model belongs to the family of strongly correlated fermion systems rather 
than that of quantum field theories.
Still, there is ample 
motivation to pursue its study. First, like continuum QCD, it confines colored
objects, and has, for $m_q \!=\! 0$, a continuous global $U(1)$ symmetry 
\be
\chi(x) \to e^{i\varepsilon(x)\alpha}\chi(x),~
\bar\chi(x) \to \bar\chi(x) e^{i\varepsilon(x)\alpha} 
~\forall ~x\,,
\label{eq:chiral}
\ee
where $\varepsilon(x) \!=\! (-1)^{\sum_4 x_\nu}$.
This is the 1-flavor version of chiral symmetry, and is spontaneously broken
in the vacuum and restored at high temperature or density.
Second, this model has been the object of analytic mean-field treatment 
since the earliest days of lattice QCD~\cite{MF_pioneer}, continuing up to 
now~\cite{Kawamoto2007,Nishida2003,Miura2008_quarkyonic}.
These approximate analytic predictions should be checked against
numerical simulations using an exact algorithm. Finally, all the obstacles for
large $\beta$ mentioned earlier can be tamed by exploiting
an alternative strategy, practical when $\beta=0$.

Instead of performing the Grassmann integral in Eq.(\ref{eq:Z}) and obtaining
a determinant which becomes a complex function of the gauge links $U$ when
$\mu\neq 0$, one integrates analytically over the gauge links first~\cite{Rossi1984}.
After this step, only colorless objects survive and propagate from one
lattice site to a neighboring site: mesons (pions), represented by unoriented ``dimers''
joining the two sites, and baryons, represented by oriented triple bonds.
Because each site hosts $N_c=3$ Grassmann fields $\chi$ and 3 fields $\bar\chi$,
a triality constraint arises: each site is attached to exactly 3 dimers, or is traversed by a baryon loop.
Thus, baryon loops are self-avoiding. After performing the Grassmann integration, the partition function becomes
a weighted sum over configurations of dimers ($n_{x,\hat\nu}=0,..,3$ dimers for each link $x,\hat\nu$) and self-avoiding baryon 
loops $C$~\cite{Karsch88}:
\be
        Z(\mu) = \displaystyle\sum_{\left\{n_{x,\hat\nu},C\right\}}
        \prod_{x,\hat\nu}\gamma^{2\delta_{\nu,4} n_{x,\hat 4}}\frac{(3-n_{x,\hat\nu})!}{n_{x,\hat\nu}!}
        ~ \prod_{C}w(C)\,,
\label{eq:Z_loop}
\ee
where a weight $w(C) \!=\! \rho(C) \gamma^{3N_{\hat{4}}(C)}\exp{\left(3k\mu a N_\tau/\gamma\right)}$ is associated with each baryon loop
$C$. Here, $N_{\hat{4}}(C)$ is the number of timelike links on the loop, $k$ is its winding number in this direction 
and $\rho(C)\!=\! \pm 1$ is a geometry-dependent sign. Thus, the weight of a configuration can be negative even when $\mu\!=\! 0$,
which seems much worse than the traditional strategy.
Yet, this sign problem can be solved by analytically resumming configurations where $C$ is a baryon loop or
a self-avoiding pion loop made of alternating single and double dimers~\cite{Karsch88}. 
After this step, the sign problem remains mild even when $\mu\!\neq\! 0$~\cite{phasediag}, so that
lattices of size $16^3\times 4$ can be simulated using the standard technique of
reweighting at all values of $\mu$, thus allowing for a reliable determination
of the full $(\mu,T)$ phase diagram. 
The final technical difficulty is to devise a Monte Carlo algorithm which
preserves the triality constraint at each site. This is achieved by the
worm algorithm~\cite{Prokof'ev2001}, adapted for strong coupling $SU(2)$ and $U(3)$
lattice theories in~\cite{Shailesh}, and readily modified here for $SU(3)$.
It produces global updates whose high efficiency does not degrade as $m_q \!\to\! 0$.

\emph{Setting the scale.} 
We first compared, at $T\!=\!\mu\!=\!0$ and for non-zero quark mass, the results of our approach with those of the traditional sampling of the 
fermion determinant by Hybrid Monte Carlo on the same $8^3\times 16$ lattice size~\cite{Kim}: there was complete agreement.
Then, setting $m_q\!=\!0$, we extracted the baryon mass $m_B$ from its Euclidean correlator $G(\tau)$,
obtaining $a m_B \!=\! -1/2 \log G(\tau+2a)/G(\tau) \!=\! 2.88(1)$, in
close agreement with mean-field~\cite{MF_pioneer} and large-$N_c$~\cite{Martin1983} predictions.
Equating our baryon mass with the real-world proton mass gives $a\!\approx\! 0.63$ fm. 
Using instead the $\Delta$ mass, which is perhaps more appropriate for our 
one-flavor model, gives $a\!\approx\! 0.46$~fm.
Alternatively, from the $\rho$ meson mass and the pion decay constant we obtain $a\!\approx\! 0.455$~fm and $a\!\approx\! 1.4$~fm, respectively. 
These different values give us an estimate of the effective coarseness of our lattice, and their dispersion gives an idea of the magnitude
of our systematic errors when comparing to real-world QCD:
our model is only a caricature of the latter.

\begin{figure}[t]
	\centerline{\includegraphics*[width=7cm,height=4.5cm]{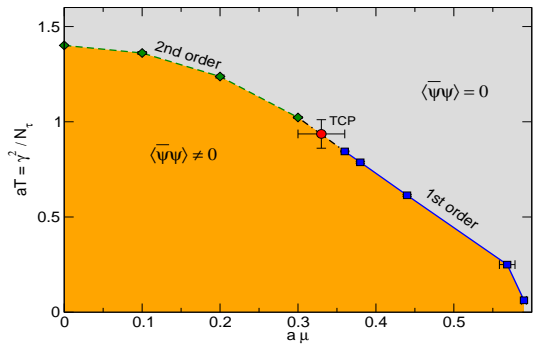}}
	\caption{$(\mu,T)$ phase diagram of 1-flavor strong coupling QCD, with massless staggered fermions ($N_\tau=4$).}
	\label{fig:phase_diagram}
\end{figure}

\emph{Phase diagram.}
We now turn to the phase diagram as a function of temperature $T$ and quark chemical potential $\mu$~\cite{phasediag}.
Since we take the quark mass to vanish, the chiral symmetry Eq.(\ref{eq:chiral}) is exact but 
spontaneously broken at small $(\mu,T)$, with order parameter $\langle \bar\psi \psi \rangle$.
When $\mu\!=\!0$, a mean-field analysis predicts symmetry restoration at $a T_c\!=\!5/3$, whereas the Monte Carlo 
study of \cite{Boyd91} on $N_\tau\!=\!4$ lattices, extrapolated to $m_q\!=\!0$, finds $a T_c \!=\! 1.41(3)$. Here, simulating directly at $m_q\!=\!0$, we assume critical exponents of the expected $3d~O(2)$ universality class, and find $a T_c \!=\! 1.319(2),1.402(3),1.417(3)$, respectively, for $N_\tau\!=\!2, 4, 6$, indicating an $N_\tau \!\to\! \infty$ limit about
$15\%$ smaller than the mean-field prediction. At $T\!=\!0$ the transition is strongly first-order, as
we will see shortly. A tricritical point separates the regimes of first- and second-order transitions.
Using finite-size scaling on $N_\tau\!=\!4$ lattices, we determine its location to be $(a \mu_{\rm TCP},a T_{\rm TCP}) \!=\! (0.33(3),0.94(7))$ 
(see Fig.~\ref{fig:phase_diagram}). This should be compared with the analytic prediction 
$(0.577,0.866)$ of~\cite{Nishida2003}. The rather large difference in $\mu_{\rm TCP}$ underlines the ${\cal O}(1/d)$ accuracy of
a mean-field treatment, and justifies {\it a posteriori} our Monte Carlo study.

In spite of the resemblance of Fig.~\ref{fig:phase_diagram} to the expected deconfinement transition in massless two-flavor QCD, 
here the two phases are both confining, with pointlike mesons and baryons, and so the phase transition is to dense, chirally symmetric, crystalline
nuclear matter.
At $T\!=\!0$ the baryon density jumps from 0 to 1, a saturation value caused by the self-avoiding
nature of the baryon loops, which itself originates from their fermion content. In physical units, this 
represents about 4 ``nucleons'' per fm$^3$, around 25 times the real-world value.

\begin{figure}[t]
 	\centerline{
	\subfigure[]{\includegraphics*[scale=0.08]{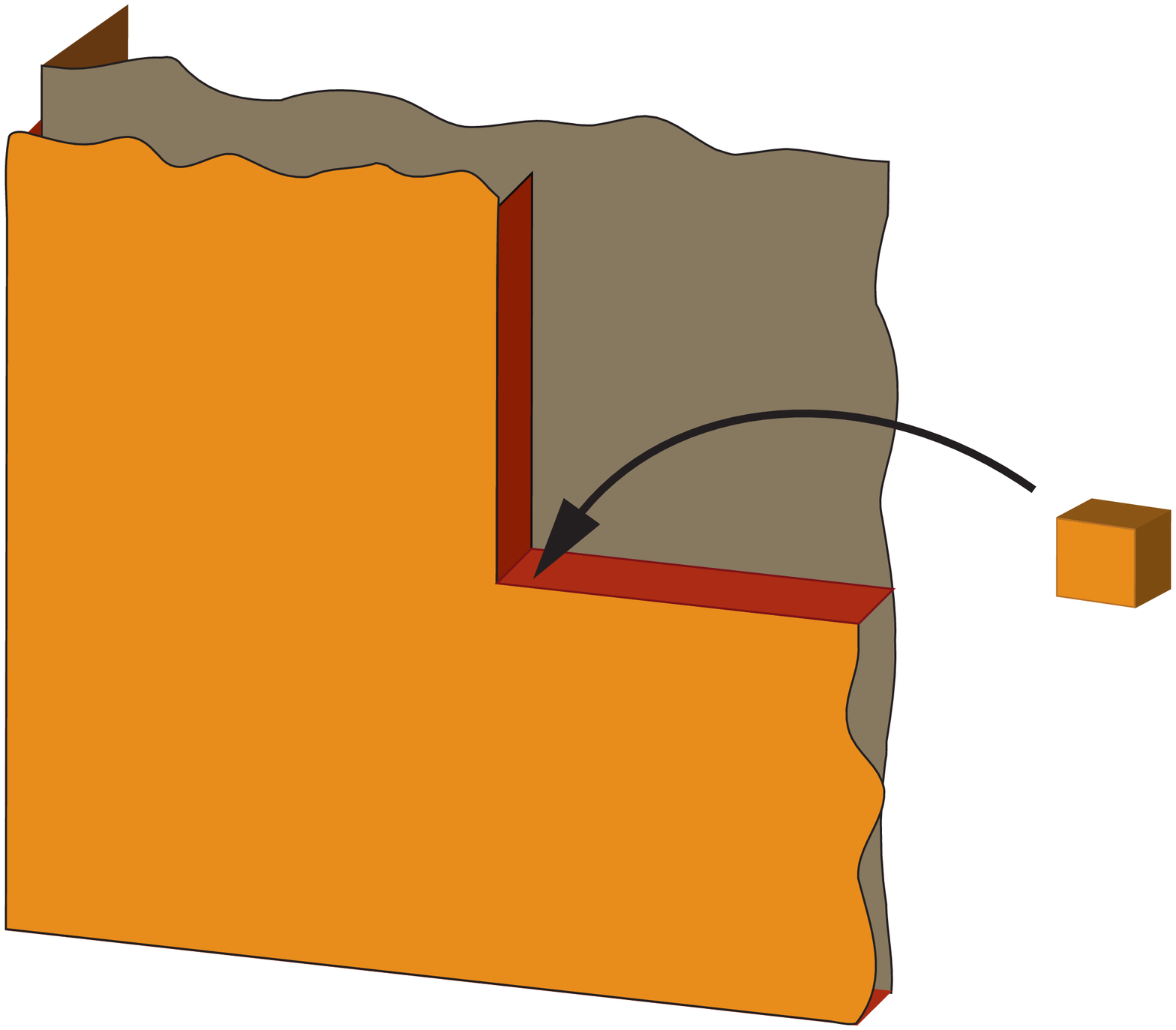}
	\label{fig:domain_wall}}
	\hspace{20pt}
	\subfigure[]{\includegraphics*[scale=0.08]{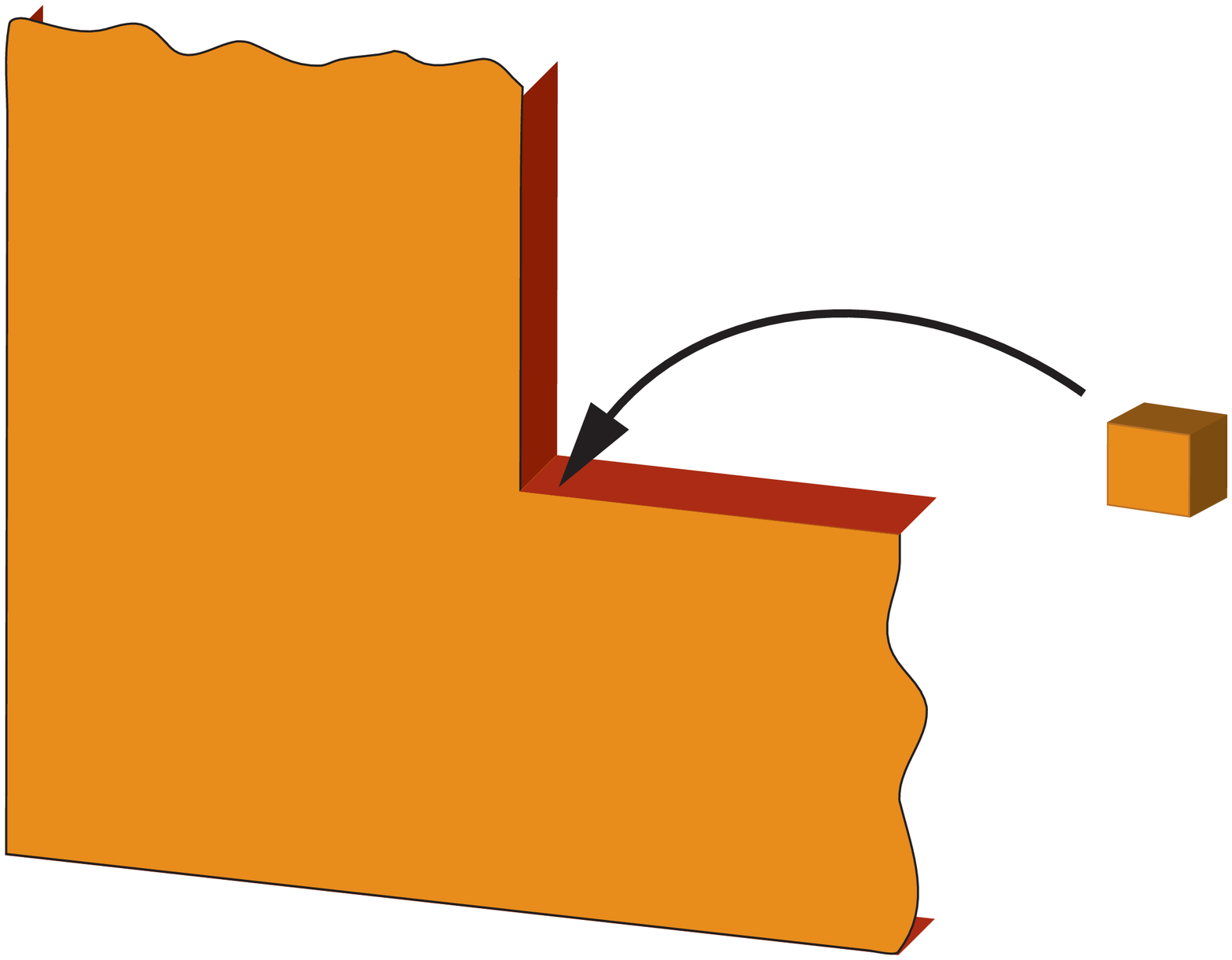}
	\label{fig:domain_wall_only}}}
	\caption{(a)~Adding a baryon to grow an additional layer of bulk nuclear matter. Each new baryon binds to 3 nearest neighbors.~(b)~Building 
	a first layer of nuclear matter inside the hadron gas, thus creating two interfaces. Each new baryon binds to 2 nearest neighbors.}	
\end{figure}

An intriguing feature of this $T\!=\!0$ transition is the value of $\mu^{\rm critical}$, which both mean-field~\cite{MF_pioneer}
and an early Monte Carlo study~\cite{Karsch88} find much smaller than the naive threshold value $m_B/3$. 
However, the ergodicity of the simulations of \cite{Karsch88} was questioned in \cite{Aloisio1999}, which was found to be justified in~\cite{Fromm:2008}.
This motivated us to redetermine $\mu^{\rm crit}(T\!=\!0)$ using an improved
method inspired by the ``snake'' algorithm~\cite{deForcrand2000}: When two phases coexist,
the free energy necessary to increase by a ``slice''
$L\times L \times a$ the volume occupied by dense nuclear matter can be decomposed into $L^2$
elementary contributions, looking generically like Fig.~\ref{fig:domain_wall}, where one additional static baryon is 
attached to 3 neighbors. We measured the free energy $\Delta F/T$ of this elementary increment on 
a large $8^3\times16$ lattice, and obtained $a \Delta F \!=\! a \mu_B^{\rm crit} \!=\! 1.78(1)$, rather close to
both mean-field predictions~\cite{MF_pioneer} and Monte Carlo extrapolations~\cite{Karsch88}, but much smaller than $a m_B$.  As already recognized in \cite{Bilic:1991}, the reason that $\mu_B^{\rm crit} < m_B$ must then be the presence of a strong nuclear attraction.

\begin{figure}[t]
	   \centerline{\includegraphics*[scale=0.4]{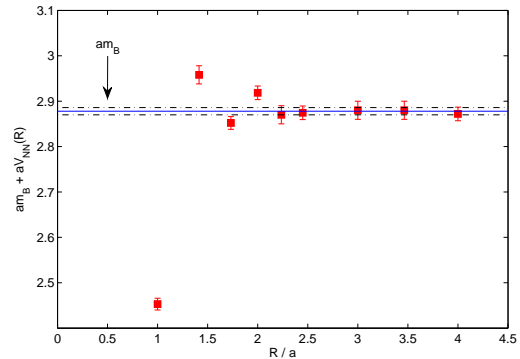}}
             \caption{Energy of a second static baryon at distance $R$ from the
first, {\it i.e.}, $(m_B + V_{\rm NN}(R))$,
             where $V_{\rm NN}(R)$ is the nuclear interaction potential.
             The horizontal band indicates the mass of an isolated baryon and 
corresponds to $V_{\rm NN}=0$.
At $R\!=\!0$ the potential is infinitely repulsive.}
	       \label{fig:V_NN}
\end{figure}

\emph{Nuclear matter.}
Since our baryons are point-like, there is no conceptual difficulty in defining the nuclear potential
$V_{\rm NN}(R)$, unlike in the real world~\cite{Beane2008}. We measured $V_{\rm NN}$ using again the snake
algorithm, this time extending little by little in Euclidean time the worldline of a second baryon at distance
$R$ from the first. The result is shown in Fig.~\ref{fig:V_NN}. Aside from the hard-core repulsion, there is indeed
a strong nearest-neighbor attraction, a slight repulsion at distance $a\sqrt{2}$, and almost no 
interaction beyond that distance. Thus, $V_{\rm NN}$ has similar features to
the real-world nuclear interaction, whose properties are commonly ascribed
to subtle competition between attractive $\sigma$ exchange and repulsive $\omega$ exchange.
The depth of the minimum $\sim\!120$~MeV and the corresponding distance $\sim\! 0.6$~fm are even quantitatively plausible. This nearest-neighbor attraction also explains {\it a posteriori} the value of $\mu_B^{\rm crit}$: each baryon
added to the dense phase binds with 3 nearest neighbors, which reduces the increase in free energy from $a m_B$ to only
$a(m_B + 3 V_{\rm NN}(a)) \approx 1.7$, consistent with 
$a\mu_B^{\rm crit}$.

Similarly, we can predict the $T\!=\!0$ surface tension
of nuclear matter: in a periodic cubic box, when building a first slice of
nuclear matter with two interfaces in the dilute phase, each new baryon binds with
only 2 nearest-neighbors (Fig.~\ref{fig:domain_wall_only}) instead of 3 in the bulk (Fig.\ref{fig:domain_wall}), thus increasing its free energy by $|V_{\rm NN}(a)|$ for an increase of $2 a^2$ in the interface area, yielding $\sigma \!\approx\! \frac{a^{-2}}{2} |V_{\rm NN}(a)|$.

\begin{figure}[h]
	\centerline{
	\subfigure[]{\includegraphics*[scale=0.42]{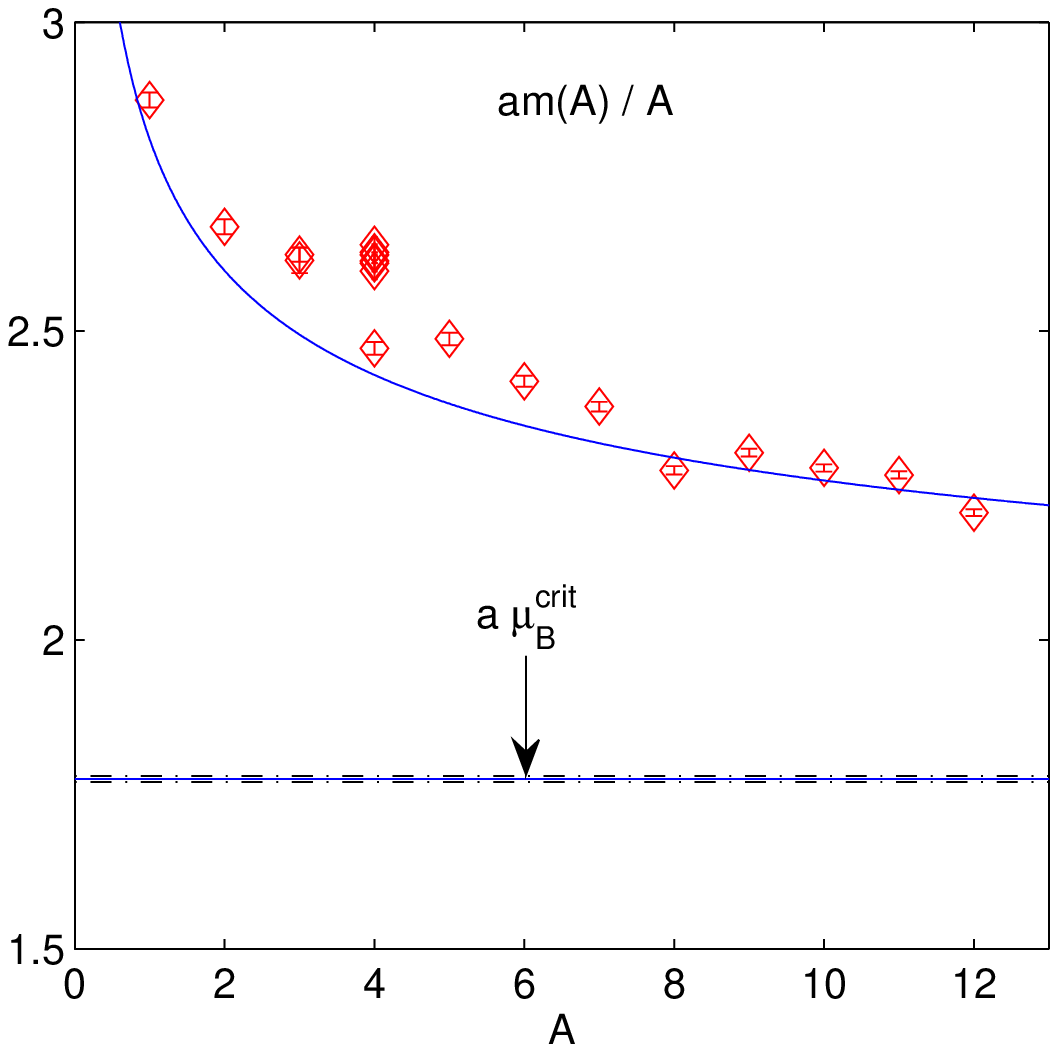}
	\label{fig:masses}}
	\subfigure[]{\includegraphics*[scale=0.15]{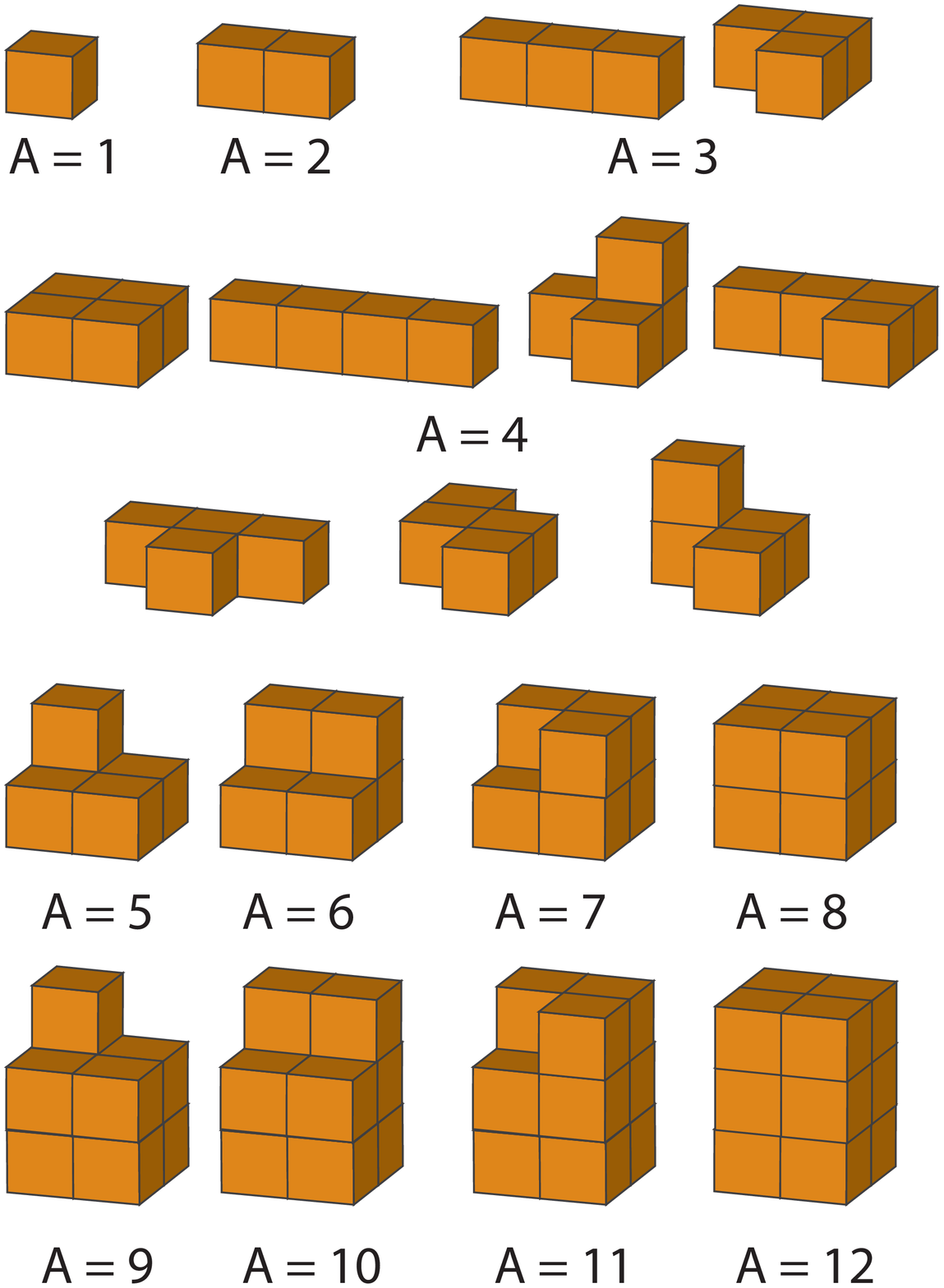}
	\label{fig:geometries}}}
	\caption{(a)~Mass per nucleon of $A\!=\!1,..,12$ nuclei. For $A\!=\!3,4$ all 
possible geometric isomers are included. The solid line 
shows the parameter-free Bethe-Weizs\"acker Eq.(\ref{eq:Bethe_Weiz}), with the surface tension $\sigma$ set
to $\frac{a^{-2}}{2} |V_{\rm NN}(a)|$.~(b)~Corresponding nuclear geometries in order of increasing mass.}	
\end{figure}

This large interface tension $\sim\!200$ MeV/fm$^2$ has an impact on the stability of ``nuclei'' of various sizes and shapes:
for a given atomic number $A$, those with a shape close to a sphere (or a cube) will have a smaller mass.
Using the same variant of the snake algorithm, we have added baryons, one by one, to form such nuclei
while measuring the successive increments in free energy.
For $A\!=\!2$ our ``deuteron'' binding energy is about
120 MeV: the real-world binding energy of $\sim\! 2$~MeV results from delicate
cancellations which do not occur in our 1-flavor model, and the binding
energy remains of the same magnitude as the depth of $V_{\rm NN}$.
For larger $A$, the resulting Fig.~\ref{fig:masses} does indeed show increased stability for
nuclei having square ($A\!=\!4$), cubic ($A\!=\!8$) or parallelepipedic ($A\!=\!12$) shapes. Other ``isomers'' with different
shapes, studied exhaustively for $A\!=\!4$ and sketched Fig.~\ref{fig:geometries},
have clearly larger masses. Moreover, the average mass per ``nucleon''
is well described by the first two (bulk and surface tension) terms of the Weizs\"acker phenomenological formula:
\be
m(A)/A = \mu_B^{\rm crit} + (36\pi)^{1/3} a^2 \sigma A^{-1/3},
\label{eq:Bethe_Weiz}
\ee
where $\sigma$ is set equal to $\frac{a^{-2}}{2} |V_{\rm NN}(a)|$ in
the figure. The next higher-order terms in this formula come from isospin and Coulomb forces, which are both absent
in our model.


\begin{figure}[t]
	\centerline{\includegraphics*[scale=0.39]{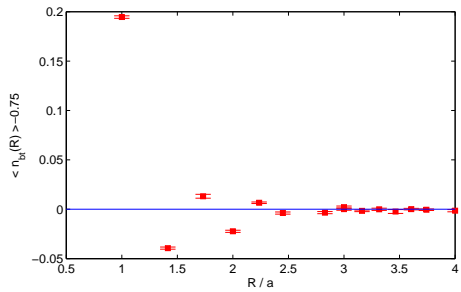}}
	\caption{Energy density of the pion cloud
as a function of the Euclidean distance to a static baryon.}
	\label{fig:n_bt}
\end{figure}

\emph{Discussion and Conclusions.}
An interesting aspect of our study is the {\em origin} of the nuclear interaction.
The nucleons are point-like and self-avoiding, so that only the hard-core repulsion is explicit.
There is no direct meson exchange in our crude model. 
In a way reminiscent of the Casimir effect between two neutral plates,
the interaction proceeds by the rearrangement of the pion bath caused by the
excluded volume of the nucleon.
This rearrangement is visible Fig.~\ref{fig:n_bt} for one nucleon: at a neighboring site, the three pion lines attached
to each site have fewer options and orient more often along the Euclidean time, which increases the pion
energy. In fact, the nucleon mass $a m_B \!\approx\! 2.88$ can be decomposed into a bare mass $3 - 3/4 \!=\! 2.25$,
which is the energy increase ``inside'' the nucleon and can be assigned to
the three valence quarks, and an energy increase $\approx\! 0.63$ in the
surrounding pion ``cloud''. When two nucleons are next to each other, the latter increase is limited to 10 nearest-neighbors
instead of $2\!\times\! 6$, which explains the attraction between them (in sign and roughly in magnitude).
This excluded volume or \emph{steric} effect is thus the origin of the nuclear potential, and ultimately of
nuclear stability, in our model. In real QCD, the pion density is not constrained as in Eq.(\ref{eq:Z_loop}).
Nevertheless, it is going to be high at temperatures $T\!\sim\! m_\pi$~\cite{Gerber1988} and one should expect
the same steric effect to enhance nuclear attraction at such temperatures. 

To summarize, in a crude model of QCD, 1-flavor lattice staggered fermions at strong coupling $\beta\!=\!0$, we have
been able to obtain the $(\mu,T)$ phase diagram and derive nuclear interactions and nuclear masses  
from first principles, uncovering a simple, but universal, steric origin of the nuclear interaction.
This model can be improved in many ways. One simple modification consists of giving a non-zero mass to
the quarks: the nuclear interaction will weaken as the pion mass is increased,
in a way which can be compared with effective field theories.
Less simple but feasible improvements include introducing isospin with
a second quark flavor, and measuring the ${\cal O}(\beta)$ correction
as done analytically in \cite{Faldt:1985,Bilic_Karsch:1991}. 
These will bring our model much closer to real QCD.

\emph{Acknowledgments}. We thank J.P.~Blaizot, S.~Chandrasekharan, N.~Kawamoto, S.~Kim, A.~Kurkela, A.~Ohnishi,
M.~Panero, J.~Rafelski, U.~Wenger and U.-J.~Wiese for discussions. The work of M.F. was supported by ETH Research Grant No. TH-07 07-2.
P.d.F. thanks KITPC, where this work was completed, for hospitality.

\end{document}